Sparse Distributed Representation, Hierarchy, Critical Periods, Metaplasticity: The Keys to Lifelong Fixed-time Learning and Best-Match Retrieval

Among the more important hallmarks of human intelligence, which any artificial general intelligence (AGI) should have, are the following.

1. It must be capable of on-line learning, including with single/few trials.
2. Memories/knowledge must be permanent over lifelong durations, safe from catastrophic forgetting. Some confabulation, i.e., semantically plausible retrieval errors, may gradually accumulate over time.
3. The time to *both*: a) learn a new item; and b) retrieve the best-matching / most relevant item(s), i.e., do similarity-based retrieval, must remain *constant* throughout the lifetime.
4. The system should never become full: it must remain able to store new information, i.e., make new permanent memories, throughout very long lifetimes.

No artificial computational system has been shown to have all these properties. Here, we describe a neuromorphic associative memory model, Sparsey, which does, in principle, possess them all. We cite prior results supporting possession of hallmarks 1 and 3. Our current primary research focus is to show possession of hallmarks 2 and 4. We sketch the argument in Sec. 2 after summarizing Sparsey in Sec. 1.

## 1. Brief Summary of Sparsey

Sparsey's most important characteristics are:

A. Information is represented using a *sparse distributed representation* (SDR) format. A single coding field consists of $Q$ WTA competitive modules (CMs), each with $K$ binary units, as shown in Figure 1a. Thus, all codes stored in the field are of the same fixed size, $Q$. Whole SDR codes are activated / deactivated in unitary fashion. Thus, the SDR, or *cell assembly*, is functionally the atomic coding unit. The SDR coding field is proposed as a model of the L2/3 population of a cortical macrocolumn, so we refer to coding fields as "macs".

B. An overall model instance is a *hierarchy* with each level consisting of an array of macs, as in Figure 1c, with local, bottom-up (U), top-down (D), and horizontal (H) connectivity.

C. Event-based *critical periods* are explicitly imposed. Specifically, learning in a mac, i.e., storing new items of information, and more specifically, storing the SDR codes assigned to represent those items, is permanently shut off when a certain fraction of all afferent synapses to the mac have been increased, i.e., when a certain *saturation* threshold of the afferent weight matrix has been reached. There is substantial evidence for critical periods in primary cortical areas [1-8] and olfactory bulb [9, 10].

D. Unit activation lifetime, or *persistence*, increases (in studies thus far, doubles) with level. Since SDRs activate / deactivate in unitary fashion, this means the persistence of whole SDR codes similarly increases with level. E.g., in Figure 1c, L1 codes persist for one frame, L2 codes for two frames, and L3 codes for four frames.

E. A large-delta Hebbian learning scheme combined with a simple metaplasticity concept. Whenever a synapse experiences a pre-post coincidence, the weight is set to the maximum possible value: weights are byte-valued, so, 127. The weight then passively decays according to its permanence (resistance to decay) value. If a second pre-post coincidence occurs within a permanence-dependent time window (typically much longer than a single sequence), the weight is reset to the max value and the permanence is increased, i.e. the rate of decay is decreased and the window length increases. Assuming that the expected time for an input reflecting a structural regularity of the world to recur is much smaller, likely exponentially smaller, than the expected time for any given random input pattern to recur, this metaplasticity scheme should preferentially embed SDRs (and chains of SDRs) reflecting the domain's structural regularities, i.e., its statistics. This permanence scheme is purely local (only requiring keeping track of time since a synapse's last weight increase) and computationally far simpler than other schemes which require repeated/continual evaluation of each weight's importance to (the constantly increasing number of) previously learned tasks/mappings. E.g., Fusi et al's Cascade model [11], Kirkpatrick et al's Elastic Weight Constraints [12], and Aljundi et al's Memory Aware Synapses [13].

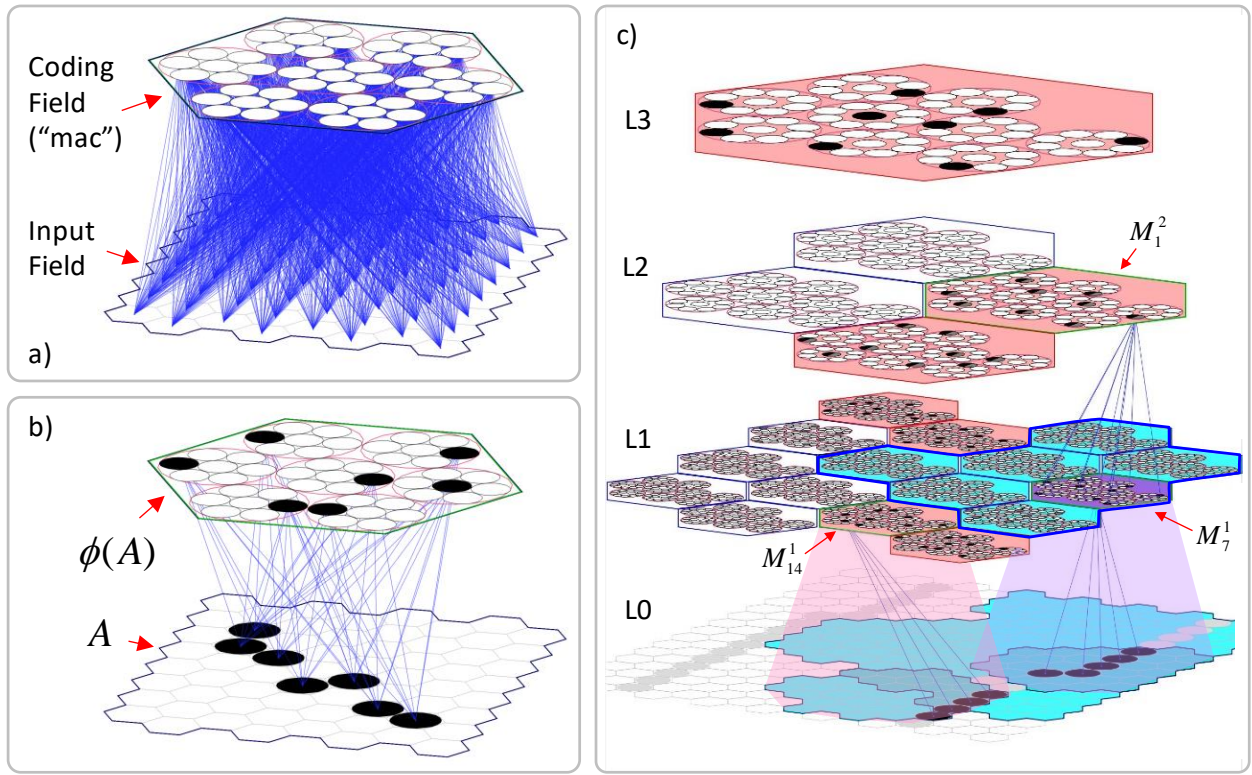

**Fig. 1.** a) An SDR coding field ("mac") consisting of $Q$=7 WTA competitive modules (CMs), each with $K$=7 units, with full U connectivity from 8x8 binary pixel input field. b) An input A, its SDR code, $\phi(A)$, consisting of $Q$ active (black) units, and the weights increased to form the association. c) Hierarchical model showing subset of active macs (rose, violet) across all levels for single input frame of a sequence. Large cyan patch at L0 is the combined L0 receptive field (RF) of the seven L1 macs (heavy blue border) comprising L2 mac $M_1^2$'s immediate (L1) RF. Smaller L0 RFs of two L1 macs also shown (emphasized by semi-transparent regions). RFs can overlap to varying extents.

Figure 2 summarizes results [14] showing that an SDR-based mac has high storage capacity and can learn arbitrary-length, complex sequences (sequences where items can repeat many times in varying context, e.g., text), with single trials, thus, addressing Hallmark 1 above. In recent work (unpublished), Sparsey achieved 90% classification accuracy on MNIST [15] (details) and 67% on Weizmann event recognition [16] (details). The accuracy is sub-SOA, but learning is extremely fast, likely much faster than SOA when adjusted for not using machine parallelism. Furthermore, it can readily be seen by inspection that Sparsey's core algorithm, the *code selection algorithm* (CSA) (see [14, 17, 18]), which is used for both learning and retrieval (recognition, inference, recall, prediction), runs in *fixed time*, i.e., the number of steps needed to both store an item and retrieve the best-matching stored item *remains constant* as the number of stored items increases, thus addressing Hallmark 3. Indeed, the essential insight of the CSA is a fixed-time method for retrieving the most similar stored item without having to explicitly compare the input (query) item to each stored item, nor to a log number of the stored items, as is the case for any tree-based method.

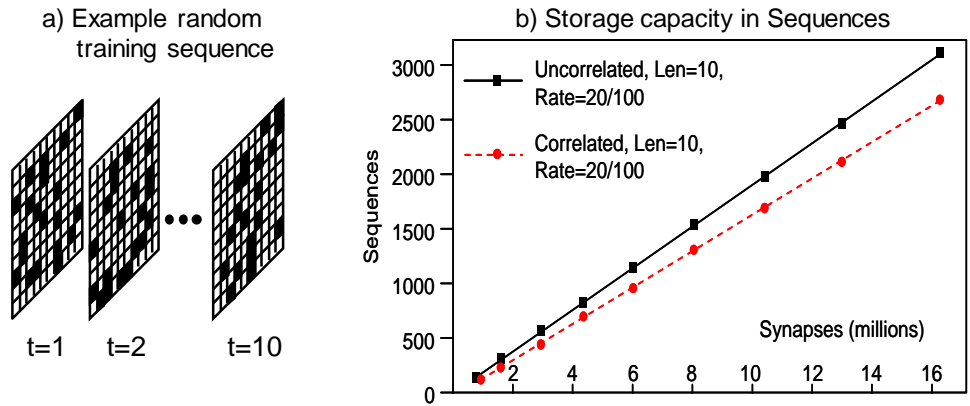

**Fig. 2:** a) Example synthetic data sequence (10 frames, 20 out of 100 randomly chosen features per frame) for testing model capacity. b) Capacity is linear in the weights. "Correlated" denotes the complex sequence case (100 frames pre-created. Sequences created by choosing from lexicon randomly with replacement).

*2. Sketch of argument for life-long learning capability without catastrophic forgetting*

We argue that the combined effects of SDR, graduated persistence, imposed critical periods, metaplasticity, and the statistics of natural inputs, cause the expected time for a mac's afferent synaptic matrices to reach saturation (i.e., for the mac to reach storage capacity) to increase quickly, likely, exponentially, with level. In practice, this means that in a system with even a few levels, e.g., ~10, as relevant to human cortex, the macs at the highest levels might never reach capacity, even over very long lifetimes operating on (or in) naturalistic domains, thus meeting Hallmark 4 above. Furthermore, since processing any single spatial input, e.g., a frame of a sequence: i) involves a single iteration over the macs from the lowest to the highest level of the hierarchy; and ii) every mac runs in fixed time; the overall hierarchical system performs learning and best-match retrieval in fixed time throughout its lifetime.

Step 1: The natural world is strongly recursively compositional in both space in time. Objects are made of parts, which are made of sub-parts, etc. Even allowing for articulation at multiple levels, this vastly constrains the space of percepts likely to be experienced compared to if all pixels of the visual field varied completely independently. Further, it has often been pointed out that visual edges carry most of the information in images/video. Consequently, Sparsey's input images (frames) are edge-filtered, binarized, and skeletonized (all cheap, local operations), as in Figure 3, which shows a few preprocessed frames from a Weizmann video. This preprocessing further greatly constrains the space of likely percepts.

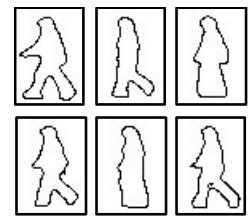

**Fig. 3**

Step 2: The human visual system is hierarchical and as suggested in Figure 1c, the receptive field (RF) size increases with level (due to cumulative fan-in/out of connections). The lowest level (L1) macs, proposed as analogs of V1 hypercolumns, have small RFs, here ~40 pixels, corresponding roughly to a diameter of 6-7 pixels. Given the above preprocessing, Sparsey implements the further policy that an L1 mac only activates if the number of active pixels in its RF is within some tight range around its RF's diameter, e.g., 6-8 pixels. This yields an input space of $C(40,6)+C(40,7)+C(40,8) \approx 10^8$. However, the combined effect of natural statistics and the preprocessing precludes the vast majority of those patterns, likely resulting in a possible input space of order tens of thousands.

Step 3: The relatively small and structurally constrained space of inputs likely to occur in a small RF, e.g., the RF of an L1 mac in Figure 1c, suggests that a relatively small basis (lexicon) of features might be sufficient to represent all future inputs to the RF with sufficient fidelity to support correct classification of the larger-scale objects/events pertinent to the RFs of macs at higher levels. That is, when an overall classification process is realized as a hierarchy of component classification processes occurring in many macs, whose RFs span many spatiotemporal scales, some substantial portion of the information as to the input's class resides in which macs are involved at each level. This decreases the needed accuracy of component classifications occurring in any individual mac, suggesting that a smaller basis—entailing a greater average difference between the mac's actual input and the best-matching stored input (basis element) to which it is mapped—may be sufficient. Fig. 4 illustrates the basic idea [here, the non-depicted higher-level recognizer's RF includes the whole field of depicted smaller-scale RFs (green hexagons)]. The presence of a plane is discernible given a very small basis set for the individual mac RFs (green hexagons), as in Fig. 4b, and arguably, even if the activated basis elements are randomly chosen (Fig. 4c). This suggests that learning can be frozen, i.e., the critical period can be terminated, in low-level macs even relatively early in the system's operational lifetime. Thus, if we factor the recognition problem into a sequence of scale-specific sub-problems (e.g., carried out at the different levels of a hierarchy), the number of samples needed to train each scale might be small and the number of samples needed overall might be exponentially smaller than for the unfactored "flat" approach, cf. Poggio and colleagues' description [19] of the hierarchical ventral stream as accomplishing this reduction of *sample complexity*. Indeed, this intuition is at the heart of graphical probability models (GPMs), such as Bayesian belief nets [20], and Hinton's Capsules model [21]. Also, note that similar considerations apply for the temporal dimension as well, i.e., natural events have strongly constrained, hierarchical temporal structures.

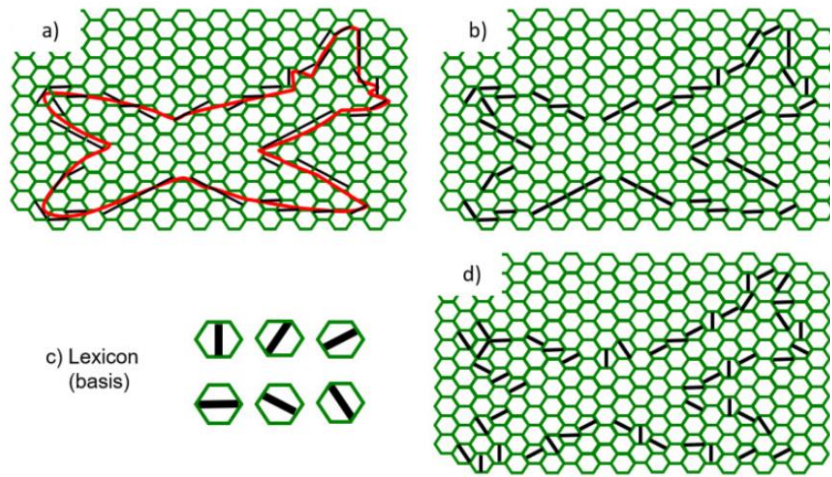

**Fig. 4**

Step 4: As also suggested in Figure 1c, a level J+1 mac's RFs consists of a patch of level J macs. Once the level J macs' bases are frozen, the space of possible inputs to level J+1 macs is further constrained. Moreover, just as an L1 mac only activates if the number of active pixels in its RF is within a tight range, so too, a level J+1 mac only activates if the number of active level J macs in its RF is within a tight range. While the space of possible raw (L0) inputs falling within the L0 RF of a level J mac increases exponentially with J, only a tiny fraction of those possible inputs ever occur. And, only a tiny fraction of those that occur, occur more than once. The permanence policy acts to let memory traces of rare events gradually fade, thus effectively reducing the rate at which higher level macs approach saturation threshold.

This is only a sketch of an argument that an SDR-based hierarchical system can retain the ability to learn new information throughout essentially unbounded lifetimes, without suffering catastrophic forgetting, and while retaining fixed response for learning and best-match retrieval. Note that catastrophic forgetting is avoided in two different ways.

1. In low-level macs, it is avoided by explicitly preventing further learning, which serves to stabilize the feature lexicons (bases) that such low-level macs will use to communicate to higher levels throughout the system's lifetime. Amongst other things, this means that lower-level macs will be computing partial invariances, i.e., invariances over their respective RFs, leaving progressively higher-level invariances to be handled at progressively higher levels, cf. [22-24].

2. In higher-level macs, it is avoided by the combined *effects* of natural statistical constraints, i.e., the aforementioned recursive part-whole structure of natural objects/events, the described preprocessing, and the explicit freezing of lower level macs' bases, Thus, at higher network levels, these principles, along with graduated persistence and metaplasticity, have the dual effects of preventing catastrophic forgetting and allowing new learning throughout the system's lifetime, thus providing a solution to Grossberg's "stability-plasticity' dilemma [25].

Prior work has shown that Sparsey possesses some of the hallmarks of biological intelligence noted at the outset. Our current research focus is to empirically validate the sketched argument, and thus the other hallmarks, through simulation.

*References*

1. Barkat, T.R., D.B. Polley, and T.K. Hensch, *A critical period for auditory thalamocortical activity". Nature Neuroscience.* Nature Neurosci., 2011. **14**(9): p. 1189-1196.
2. Friedmann, N. and D. Rusou, *Critical period for first language: the crucial role of language input during the first year of life.* Current Opinion in Neurobiology, 2015. **35**: p. 27-34.
3. Erzurumlu, R.S. and P. Gaspar, *Development and Critical Period Plasticity of the Barrel Cortex.* The European Journal of Neuroscience, 2012. **35**(10): p. 1540-1553.
4. Muir, D.W. and D.E. Mitchell, *Behavioral deficits in cats following early selected visual exposure to contours of a single orientation.* Brain Research, 1975. **85**(3): p. 459-477.
5. Hubel, D.H. and T.N. Wiesel, *The period of susceptibility to the physiological effects of unilateral eye closure in kittens.* The Journal of Physiology, 1970. **206**(2): p. 419-436.
6. Blakemore, C. and R.C. Van Sluyters, *Reversal of the physiological effects of monocular deprivation in kittens: further evidence for a sensitive period.* The Journal of Physiology, 1974. **237**(1): p. 195-216.
7. Pettigrew, J.D. and R.D. Freeman, *Visual Experience without Lines: Effect on Developing Cortical Neurons.* Science, 1973. **182**(4112): p. 599-601.
8. Daw, N.W. and H.J. Wyatt, *Kittens reared in a unidirectional environment: evidence for a critical period.* The Journal of Physiology, 1976. **257**(1): p. 155-170.
9. Cheetham, C.E. and L. Belluscio, *An Olfactory Critical Period.* Science, 2014. **344**(6180): p. 157-158.
10. Poo, C. and J.S. Isaacson, *An Early Critical Period for Long-Term Plasticity and Structural Modification of Sensory Synapses in Olfactory Cortex.* J. Neurosci., 2007. **27**(28): p. 7553-7558.
11. Fusi, S., P.J. Drew, and L.F. Abbott, *Cascade Models of Synaptically Stored Memories.* Neuron, 2005. **45**(4): p. 599-611.
12. Kirkpatrick, J., et al., *Overcoming catastrophic forgetting in neural networks.* PNAS, 2017. **114**(13): p. 3521-3526.
13. Aljundi, R., et al. *Memory Aware Synapses: Learning what (not) to forget*. ArXiv e-prints, 2017. **1711**.
14. Rinkus, G., *A Combinatorial Neural Network Exhibiting Episodic and Semantic Memory Properties for Spatio-Temporal Patterns*, in *Cognitive & Neural Systems*. 1996, Boston U.: Boston.
15. LeCun, Y., et al., *Gradient-based learning applied to document recognition.* Proc. IEEE, 1998. **86**: p. 2278–2324.
16. Gorelick, L., et al., *Actions as Space-Time Shapes.* IEEE Trans. PAMI, 2007. **29**(12): p. 2247-2253.
17. Rinkus, G., *A cortical sparse distributed coding model linking mini- and macrocolumn-scale functionality.* Frontiers in Neuroanatomy, 2010. **4**.

18. Rinkus, G.J., *Sparsey™: event recognition via deep hierarchical sparse distributed codes.* Frontiers in Computational Neuroscience, 2014. **8**(160).
19. Poggio, T., et al., *The computational magic of the ventral stream: sketch of a theory (and why some deep architectures work).* 2012, MIT CSAIL.
20. Pearl, J., *Probabilistic Reasoning in Intelligent Systems: Networks of Plausible Inference* 1988, San Mateo, CA: Morgan Kaufmann.
21. Sabour, S., N. Frosst, and G. E Hinton *Dynamic Routing Between Capsules.* ArXiv e-prints, 2017. **1710**.
22. Rust, N.C. and J.J. DiCarlo, *Selectivity and Tolerance ("Invariance") Both Increase as Visual Information Propagates from Cortical Area V4 to IT.* The Journal of Neuroscience, 2010. **30**(39): p. 12978-12995.
23. Zoccolan, D., et al., *Trade-Off between Object Selectivity and Tolerance in Monkey Inferotemporal Cortex.* J. Neurosci., 2007. **27**(45): p. 12292-12307.
24. Fusi, S., E.K. Miller, and M. Rigotti, *Why neurons mix: high dimensionality for higher cognition.* Current Opinion in Neurobiology, 2016. **37**: p. 66-74.
25. Grossberg, S., *How does a brain build a cognitive code?* Psychological Review, 1980. **87**(1): p. 1-51.